# Lessons Learned from MammoGrid for Integrated Biomedical Solutions

R.H.McClatchey[1], D. Manset[2] & A.E.Solomonides[1]

*1 CCCS, UWE Bristol, Frenchay Campus, Bristol BS16 1QY, UK*
*2 Maat GKnowledge, Toledo, Spain*

*{tony.solomonides / richard.mcclatchey}@uwe.ac.uk, dmanset@Maat-G.com*

## Abstract

*This paper presents an overview of the MammoGrid project and some of its achievements. In terms of the global grid project, and European research in particular, the project has successfully demonstrated the capacity of a grid-based system to support effective collaboration between physicians, including handling and querying image databases, as well as using grid services, such as image standardization and Computer-Aided Detection (CADe) of suspect or indicative features. In terms of scientific results, in radiology, there have been significant epidemiological findings in the assessment of breast density as a risk factor, but the results for CADe are less clear-cut. Finally, the foundations of a technology transfer process to establish a working "MammoGrid plus" system in Spain through the company Maat GKnowledge and the collaboration of CIEMAT and hospitals in Extremadura.*

## 1. Background

Breast cancer is arguably the most pressing threat to women's health. For example, in the UK, more than one in four female cancers occur in the breast and these account for 18% of deaths from cancer in women. Coupled with the statistic that about one in four deaths in general are due to cancer, this suggests that nearly 5% of female deaths are due to breast cancer. While risk of breast cancer to age 50 is 1 in 50, risk to age 70 increases to 1 in 15 and lifetime risk has been calculated as 1 in 9. The problem of breast cancer is best illustrated through comparison with lung cancer which also accounted for 18% of female cancer deaths in 1999. In recent years, almost three times as many women have been diagnosed with breast cancer as with lung cancer. However, the five year survival rate from lung cancer stands at 5%, while the breast cancer figure is 73%. This is testament to the effectiveness of modern treatments, provided breast cancer is diagnosed sufficiently early. These statistics are echoed in other countries. The lifetime risk of breast cancer in the USA has been estimated as 1 in 8. Here also incidence has increased but mortality decreased in the past twenty years. There is evidence that stress and diet may also be correlated with breast cancer, although the underlying biochemistry is not fully understood. For example, twenty years ago breast cancer was almost unknown in Japan but its incidence now approaches Western levels. (For a world-wide picture, see [1].) This may be associated with major changes that have occurred in Japan in this period, e.g. the increased consumption of high lipid foods and higher levels of stress with more women working. Finally, we note that a small number of men, e.g., about 100 in the UK and 400 in the US, also die annually of breast cancer.

The statistics of breast cancer diagnosis and survival appear to be a powerful argument in favour of a universal screening programme. However, a number of issues of efficacy and cost effectiveness limit the scope of most screening programmes. The method of choice in breast cancer screening is mammography (breast X-ray), although self-examination and clinical palpation are also used; for precise location of lesions and 'staging' (establishing how advanced the disease is) ultrasound and MRI may be used. A significant difficulty lies in the typical composition of the female breast, which changes dramatically over the lifetime of a woman, with the most drastic change taking place around the menopause. In younger women, the breast consists of around 80% glandular tissue which is dense and largely X-ray opaque. The remaining 20% is mainly fat. In the years leading up to the menopause, this ratio is typically reversed. Thus in women under 50, signs of malignancy are far more difficult to discern in mammograms than they are in post-menopausal women. Consequently, most screening programmes, including the UK's, only apply to women over 50. [2]

Given the decision to institute a screening programme, apart from deciding the subject age-group, another vital question is that of frequency of screening. Based on contemporary thinking and practice elsewhere in Europe, the UK programme was set up to screen every three years; however, by the time this was inaugurated, further research in Sweden suggested that screening should take place more often. Indeed, one of the problems the UK programme has faced has been the phenomenon of so-called 'interval cancers', cancers which develop substantially in the three-year interval between screens.

A further question which received a conservative answer in the UK was how many images of each breast should be taken. It is generally believed that two images of each breast are ideal: a cranio-caudal (CC) image, vertically downwards through the breast, and a medio-lateral oblique (MLO) image, at 45° from shoulder to opposite hip, taking in part of the pectoral muscle. However, in view of the resources needed to take and interpret four images, it was decided that two CC images should suffice.

The increasing use of electronic formats for radiological images, including mammography, together with the fast, secure transmission of images and patient data, potentially enables many hospitals and imaging centres throughout Europe to be linked together to form a single grid-based "virtual organization". It is not yet precisely understood what advantages might accrue to radiologists working in such virtual organizations, as the technological possibilities are co-evolving with an appreciation of potential uses; but one that is generally agreed is the creation of huge "federated" databases of mammograms, which appear to the user to be a single database but are in fact retained and curated in the centres that generated them. Each image in such a database would have linked to it a potentially large and expandable set of relevant information, known as metadata, about the woman whose mammogram it is. This might comprise: patient age, exogenous hormone exposure, family and clinical history; information known to correspond to risk factors (eg diet, parity, breast density); as well as image acquisition parameters, including breast compression and exposure data which affect image appearance. Levels of access to the images and metadata in the database would vary according to the "certificated rights" of the user: healthcare professionals might have access to essentially all of it, whereas epidemiologists, researchers, government officials would have more limited access, protecting patient privacy and in accordance with European legislation.

The Fifth Framework EU-funded MammoGrid project [3] aimed to apply the grid concept to mammography, including services for the standardization of mammograms, computer-aided detection (CADe) of salient features, especially masses and 'microcalcifications', quality control of imaging, and epidemiological research including broader aspects of patient data. In doing so, it attempted to create a paradigm for practical, grid-based healthcare-oriented projects, particularly those which rely on imaging. There are, however, a number of factors that make patient management based on medical images particularly challenging. Often very large quantities of data, with complex structures, are involved (such as 3-D images, time sequences, multiple imaging protocols etc.). Also, clinicians rarely analyse single images in isolation but rather in a series or in the context of metadata. Metadata that may be required are clinically relevant factors such as patient age, exogenous hormone exposure, family and clinical history; for the population, natural anatomical and physiological variations; and for the technology, image acquisition parameters, including breast compression and exposure data.

Thus any database of images developed at a single site may not contain enough exemplars in response to any given query to be statistically significant. Overcoming this problem implies constructing a very large, federated database, which can transcend national boundaries. However this necessitates specialist image processing algorithms – i.e., computationally heavy tasks operating on large files of images – which in turn place significant requirements on storage space, CPU power and/or network bandwidth on all participating hospitals, unless appropriate sharing of computing resources is arranged. Realising such a geographically distributed database therefore necessitates a grid infrastructure, and the construction of a prototype model which would push emerging grid technology to its limits.

The MammoGrid project was carried out between mid 2002 and the end of 2005 and involved hospitals and medical imaging experts and academics in the UK, Italy and Switzerland with experience of implementing grid-based database solutions. A key deliverable of the project was a prototype software infrastructure based on an open-source grid 'middleware' (i.e. software that enables an underlying grid infrastructure to host domain applications) and a service-oriented database management system that is capable of managing federated mammogram databases distributed across Europe. The proposed solution was a medical information infrastructure delivered on a service-based, grid-aware framework, encompassing geographical regions of varying clinical protocols and diagnostic procedures, as well as lifestyles and dietary patterns. The prototype allows, among other things, mammogram data mining, diverse and complex epidemiological studies, statistical and (CADe) analyses, and the deployment of versions of the image standardization software. It was the intention of MammoGrid to get rapid feedback from a real clinical community about the use of such a simple grid platform to inform the next generation of grid projects in healthcare.

The clinical work packages encompassed in MammoGrid prototypes address three selected clinical problems:

1. Quality control: the effect on clinical mammography of image variability due to differences in acquisition parameters and processing algorithms;
2. Epidemiological studies: the effects of population variability, regional differences such as diet or body habitus and the relationship to mammographic density (a potential biomarker of breast cancer) which may be affected by such factors;
3. Support for radiologists, in the form of tele-collaboration, second opinion, training and quality control of images.

Other initiatives against which MammoGrid may be compared include: the eDiamond [4] project in the UK, and the NDMA [5] project in the US. The MammoGrid approach shares many similarities with these projects, but in the case of the NDMA project its database is implemented in IBM's DB2 on a single server. The MammoGrid project federates multiple (potentially heterogeneous) databases as its data store(s). The Italian INFN project GP-CALMA [6] has focused on a grid implementation of tumour detection algorithms to provide clinicians with a working mammogram examination tool. MammoGrid uses aspects of the CALMA project in its computer-aided detection of microcalcifications.

More recent grid-based research includes the BIRN [7] project in the US, which is enabling large-scale collaborations in biomedical science by utilizing the capabilities of emerging grid technologies. BIRN provides federated medical data, which enables a software 'fabric' for seamless and secure federation of data across the network and facilitates the collaborative use of domain tools and flexible processing/analysis frameworks for the study of Alzheimer's disease. The INFOGENMED initiative [8] has given the lead to projects in moving from genomic information to individualized healthcare using data distributed across Europe.

From the outset, the MammoGrid project posted its objectives in terms of the promised radiological and epidemiological applications, but not in terms of new grid technology. Its technology attitude has largely been one of re-use, not invention or development; only where required functionality was missing was there a need to implement new grid services. An information infrastructure to integrate multiple mammogram databases is clearly needed to enable clinicians to develop new common, collaborative and co-operative approaches to the analysis of mammographic images as is evident by the clinical evaluation that took place towards the end of the MammoGrid project.

## 2. Results from MammoGrid

The MammoGrid project has recently delivered its final proof-of-concept prototype enabling clinicians to store digitized mammograms along with appropriately anonymized patient metadata; the prototype provides controlled access to mammograms both locally and remotely stored. A typical database comprising several thousand mammograms has been created for user tests of clinicians' queries. The prototype comprises:

- a high-quality clinician visualization workstation (used for data acquisition and inspection);
- an imaging standard-compliant interface to a set of medical services (annotation, security, image analysis, data storage and querying services) residing on a so-called 'Grid-box'; and
- secure access to a network of other Grid-boxes connected through Grids middleware.

To allow evaluation of the final prototype at the clinical sites, a MammoGrid Virtual Organization (MGVO) was established and deployed (see figure 1). The MGVO is composed of three mammography centres – Addenbrookes Hospital, Udine Hospital, and Oxford University. These centres are autonomous and independent of each other with respect to their local data management and ownership. The Addenbrookes and Udine hospitals have locally managed databases of mammograms, with several thousand cases between them. As part of the MGVO, registered clinicians have access to (suitably anonymized) mammograms, results, diagnosis and imaging software from other centres. Access is coordinated by the MGVO central node at CERN.

The adopted grid implementation is the ALICE Environment (AliEn) [9] component of the EGEE-gLite middleware [10], the grid middleware of the EU-funded EGEE project [11]. The service-oriented approach adopted in MammoGrid permits the interconnection of communicating entities, called services, which provide capabilities through exchange of messages. The services are 'orchestrated' in terms of service interactions: how services are discovered, how they are invoked, what may be invoked, the sequence of service invocations, and who can execute them.

Evaluation of the MammoGrid prototype took place using the MGVO over a set of clinical work packages performed by senior radiologists at the two hospitals. The MammoGrid Virtual Organization encompassed data accessible to the radiologists at the hospitals, as well as at Oxford University and CERN. The evaluation comprised the qualitative and, where possible, quantitative assessment of the use-cases captured during the requirements analysis phase of the project.

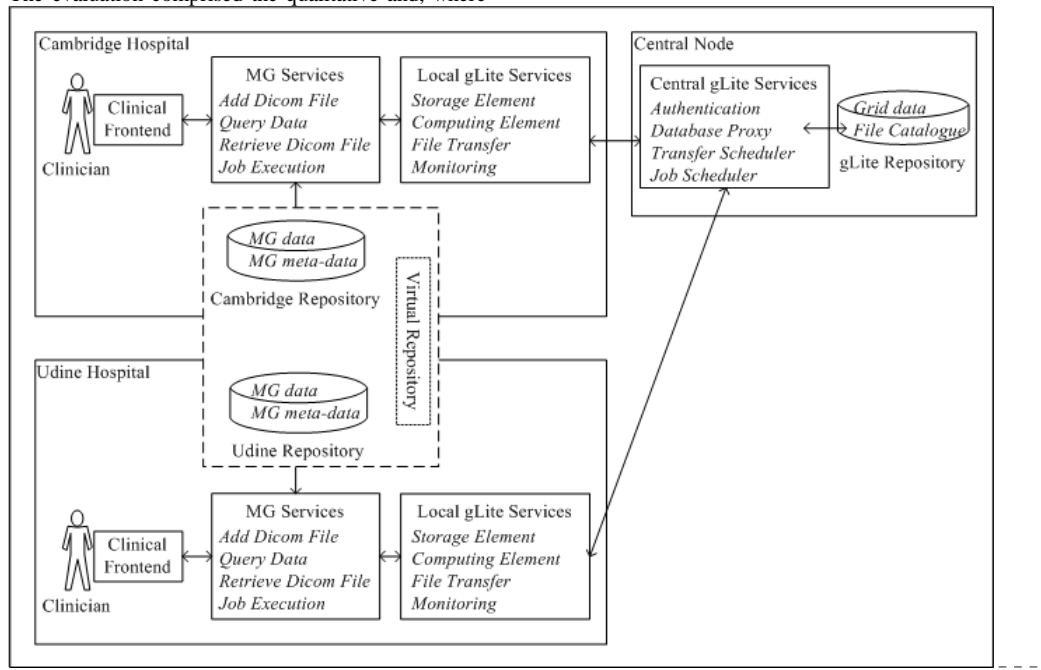

Figure 1: The MammoGrid Virtual Organisation

During the evaluation the radiologists have been able to view raw image data from each others' hospitals and have been able to second read grid-resident mammograms and to separately annotate the images for combined diagnosis. This has demonstrated the viability of distributed image analysis using the grid and shown considerable promise for future health-based grid applications. Despite the anticipated performance limitations that existing grid software imposes on the system usage, the clinicians have been able to discover new ways to collaborate using the virtual organization. These include the ability to perform queries over a virtual repository spanning data held in Addenbrookes and Udine.

Following the 'Perform Radiological Analysis' use-case scenario, identified during the project's requirements analysis phase, clinicians define their mammogram analysis in terms of queries they wish to be resolved across the federation of data repositories. Queries can be categorized into simple queries (mainly against associated data stored in the database as simple attributes) and complex queries which require derived data to be interrogated or an algorithm to be executed on a (sub-)set of distributed images. The important aspect is that image and data distribution are transparent for radiologists so queries are formulated and executed as if these images were locally resident. Queries are executed at the location where the relevant data resides, i.e. sub-queries are moved to the data, rather than large quantities of data being moved to the clinician, which can be prohibitively expensive given the quantities of data.

Clinicians continue in the process of scanning and annotating cases contributing to several ongoing medical studies. These include (1) Cancers versus control study: breast density study using the standard mammogram form (SMF™) standard [12] (2) Dose/density study: exploring the relationship between mammographic density, age, breast size and radiation dose (3) CADe and validation of SMF in association with CADe. These studies continue to show how health professionals can work together without co-locating. Importantly the collaborative approach pursued in MammoGrid has already identified new ways in which clinicians can work together using a common grid-based repository which were hitherto not possible. For example, the use of the SMF algorithm on data supported by MammoGrid and accessible to radiologists in Cambridge and Udine for the purposes of joint mammogram analysis has directly led to results recently submitted to European Radiology [13].

In summary, during the final months of the project the clinicians have been evaluating the MammoGrid prototype across two applications. First, the project has facilitated the use of the SMF software to measure breast density. The clinical project, designed jointly by Cambridge and Udine, explored the relationship between mammographic density, age, breast size, and radiation dose. In this project, breast density has been measured by SMF and compared with standard methods of visual assessment. Heights, weights, and mass indicators are used in an international comparison, although a richer dataset would be needed to study effects of lifestyle factors such as diet or HRT use between the two national populations. Second, the University of Udine led a project to validate the use of SMF in association with CADe from the CALMA project. Cancers and benign lesions have been supplied from the clinical services of Udine and Cambridge to provide the benchmarking and the set of test cases. Cancer cases will include women whose unaffected breast will serve the density study to provide cases for the CADe analysis from the affected side mammogram. MammoGrid has demonstrated that these new forms of clinical collaboration can be supported using the grid [14].

This is expected to result in improved research quality as well as improved citizen access to the latest healthcare technologies.

## 3. Technology Transfer

By using grid computing, the MammoGrid system allows hospitals, healthcare workers and researchers to share data and resources, while benefiting from an augmented overall infrastructure. It supports effective co-working, such as obtaining second opinion, provides the means for powerful comparative analysis of mammograms and opens the door to novel, broad-based statistical analysis of the incidence and forms of breast cancer. Through the MammoGrid project, partners have developed a strong collaboration between radiologists active in breast cancer research and academic computer scientists with expertise in the applications of grid computing. The success of the project has led to interest from outside companies and hospitals, with one Spanish company, Maat GKnowledge, looking to deploy a commercial variant of the system in three hospitals of the Extremadura region in Spain. Maat GKnowledge aims to provide doctors with the ability to verify test results, to obtain a second opinion and to make use of the clinical experience acquired by the hospitals involved in the project. They then aim to scale the system up and to expand it to other areas of Spain and then Europe. With the inclusion of new hospitals, the database will increase in coverage and the knowledge will increase in relevance and accuracy, enabling larger and more refined epidemiological studies. Therefore, clinicians will be provided with a significant data set to serve better their investigations in the domain of cancer prevention, prediction and diagnoses. This will result in improved research quality as well as improved citizen access to the latest healthcare technologies.

The MammoGrid system prototype has been at the leading edge of the healthgrid revolution and implements for the first time such a solution for mammogram acquisition and manipulation. The resulting application has reached a high level of complexity which now requires continued partnership between academics, clinicians and industry to provide the necessary technology transfer and to enable real commercialisation. In this context, the MammoGrid Technology Transfer and Innovation eXchange (MaTTrIX) project has been proposed as an ideal means to transfer the project knowledge and expertise from the research to the commercial domain, to make its innovation available to the company Maat GKnowledge in Spain, to carry forward research findings to radiological practice and to reinforce existing partnerships between networks of clinicians and technologists. To achieve these objectives, the intention is to introduce a service which exploits the findings of the MammoGrid project in practice, while the researchers also learn from the application of their ideas in a real environment. This will create a two-way innovation flow between the academic and commercial worlds, building on existing synergies and collaborations, and improving the overall viability and commercialisation of the MammoGrid system software.

To this end, the host organisation Maat GKnowledge, the University of the West of England, Bristol, CIEMAT (Centro de Investigaciones Energéticas, Medioambientales y Tecnológicas), Spain, and CERN together with clinical partners at the University hospitals in Cambridge and Udine and at the Hospitals Infanta Cristina, Merida and San Benito in Extremadura are initiating a transfer of knowledge, research competences and technologies to enable the company to take over future development of the MammoGrid system. This will enable the training of existing Maat GKnowledge staff in MammoGrid technologies and the acquisition of the technology and know-how for commercialisation. This transfer would constitute the mechanism for the creation and development of a durable technology and knowledge transfer partnership.

The MaTTrIX technology transfer approach relies on two main cornerstones:

- to promote the mobility of key experienced researchers, to absorb, expand and disseminate the knowledge for the MammoGrid system to evolve gradually into a commercial offering, making it available for healthcare across the European Research Area (ERA) as a viable and clinically assessed solution; and

- to create and to develop a strategic and durable partnership between new hospitals and partners of the MammoGrid project, providing a sound foundation for a network of excellence in European research.

Having paved the way for potential knowledge discovery in the understanding of breast cancer, the MammoGrid project has also determined new important research paths for better cancer prediction and diagnosis. The use of a standard format for mammogram images (SMF) and its outcome in the epidemiological investigations, has demonstrated the relevance of new grid-based clinical studies, which may lead to major advances in cancer prediction. In addition, while most computer-aided detection (CADe) systems process raw and noisy data at the price of accuracy and quality, the implemented solutions in MammoGrid have indicated the valuable joint use of SMF and CADe tools to improve automated cancer diagnosis. Considering the IT contribution in MammoGrid, the distributed technologies used in the project combined with the recent clinical feedback, obtained from the assessment of the proof-of-concept prototype, have highlighted the importance of offering a collaborative platform for healthcare. Not only would such a solution demonstrate the benefits of clinical second opinion, but it would also help in reducing the information infrastructure costs by enabling heterogeneous and scalable resource sharing, resulting in an improved system access even to less favoured regions and countries in Europe.

Emphasising this last point the Hospitals of Infanta Cristina, Merida and San Benito in Extremadura in Spain, will obtain access to the MammoGrid system and expertise at a reduced cost, the infrastructure being already in place. While accessing the latest technologies and software, these hospitals will share and enrich their clinical experience by interacting with other trained clinicians from the university hospitals in Cambridge and Udine to obtain expertise in the use of SMF and CADe. Academic computer scientists will continue to analyse the ways in which new functionality is exploited by radiologists in their protocols and workflows, maintaining the design of the system under review. This new partnership will result in improved processes locally and also in refined clinical knowledge being made available in a Europe-wide production reference database, enabling new clinical studies in the spirit of improving cancer prediction and detection. It is hoped that this collaboration will provide a significant exemplar for the European Research Area.

The MammoGrid project has deployed its first prototype and has performed the first phase of in-house tests, in which a representative set of mammograms have been tested between sites in the UK, Switzerland and Italy. In the next phase of testing, clinicians will be closely involved in performing tests and their feedback will be reflectively utilised in improving the applicability and performance of the system. In its first two years, the MammoGrid project has faced interesting challenges originating from the interplay between medical and computer sciences, and has witnessed the excitement of the user community whose expectations from a new paradigm are understandably high. As the

MammoGrid project moves into its final implementation and testing phase, further challenges are anticipated. In conclusion, this paper has outlined the MammoGrid application's deployment strategy and experiences. Also, it outlines the strategy being adopted for migration to the new lightweight middleware called gLite.

Whereas the MammoGrid system was meant to be grid middleware independent in its latest releases, the MammoGrid Plus software will be based on AliEn combined with Maat's proprietary "G" database management system. Indeed, MammoGrid Plus being a commercial product, the objective is not to make it as generic as possible to other middleware, but rather to acquire the expertise on a particular one. In a commercial context, this will highly facilitate the maintenance and issues faced in the usage of the system. Also, without an established standard for grid middleware, it is not commercially viable to open the system to several different middleware prototypes. However, it should be noted that the philosophy behind the integration of the "G" database management system as a junction layer between domain services and grid services will provide the necessary abstraction in case of a middleware replacement. AliEn functionalities will be accessible as if querying/populating a regular database, the latter acting directly on the middleware data.

Despite this fact, the MammoGrid Plus software will comply with the main design principles of MammoGrid in addressing the improvement (and enrichment if required) of the metadata database infrastructure. To do so, it will provide a proper distributed DMS layer following the metadata schemas designed in MammoGrid.

Based on Maat's recent advances in collaboration with CERN, the grid middleware used for the MammoGrid Plus product will be a modified AliEn lightweight grid middleware featured with the "G" DBMS as the basis of its File Catalogue. Aspects of the File Catalogue in AliEn have been well suited to a research environment, but are not appropriate in a commercial context where the emphasis is mainly on performance.

By leveraging middleware performance and scalability, the MammoGrid Plus product will be ready for intensive use in production environments. In addition, using the "G" system, the overall security and privacy of the software will be improved at the functionality level. Indeed, one of the weaknesses of AliEn is its poor capability of representing user's authorization/roles. Using AliEn, it is not possible to obtain a fine-grained role definition for the users of the system, the middleware only distinguishing between normal users and admin users. The "G" platform, allows for the definition of a complete set of roles and authorization accesses. Thus, in the MammoGrid Plus software it will be possible to extend the scope of the system by distinguishing different clinicians' roles in a given virtual organization, opening the room for new functionalities where each user will obtain a customized access to the software offer.

## 4. Conclusions

MammoGrid is one of several current projects that aim to harness recent technological advances to achieve the goal of complex data storage in support of medical applications. The approaches vary widely, but at least two projects, e-Diamond in the UK and MammoGrid in Europe, have adopted the grid as their platform of choice for the delivery of improved quality control, more accurate diagnosis and statistically significant epidemiology. Since breast screening in the UK and in Italy has been based on film, mammograms have had to be digitised for use in both e-Diamond and MammoGrid. By contrast, the NDMA project in the United States has opted for centralised storage of direct digital mammograms. The next step for MammoGrid, its application in the Spanish region of Extremadura, will be based on both film and direct digital mammography.

The central feature of the MammoGrid project is a geographically distributed, grid-based database of standardised images and associated patient data. The novelty of the MammoGrid approach lies in the application of grid technology and in the provision of data and tools, which enable radiologists to compare new mammograms with existing ones on the grid database, allowing them to make comparative diagnoses as well as judgements about quality. In the longer term, as the potential of the database is to be populated with provenance-controlled, reliable data from across Europe, with the prospect of statistically robust epidemiology that allows analysis of 'lifestyle' factors, including, e.g., diet, exercise and exogenous hormone use. And, hence the grid would also be suitable for storing genetic or pathological image information.

The project has attracted attention as a paradigm for grid-based radiology and imaging applications. While it has not yet solved all problems, the project has established an approach and a prototype platform sharing medical data, especially images, across a grid. In loose collaboration with a number of other European medical grid projects, it is addressing the issues of informed consent and ethical approval, data protection, compliance with institutional, national and European regulations, and security. In conclusion, the MammoGrid project may be considered as a major advance in bridging the gap between the grid as an advanced distributing computing infrastructure and the medical domain and therefore should enable further grid-based projects to benefit from both its main lessons and its results.

## 5. Acknowledgements

The authors thank the European Commission and their institutes for support and acknowledge the contribution of the following MammoGrid collaboration members: Sir Michael Brady and Chris Tromans (University of Oxford); Predrag Buncic, Pablo Saiz (CERN/AliEn), Martin Cordell, Tom Reading and Ralph Highnam (Mirada); Piernicola Oliva (University of Sassari); Evelina Fantacci and Alessandra Retico (University of Pisa); Tamas Hauer, Dmitry Rogulin and Waseem Hassan (University of the West of England/CERN). The collaboration of the MammoGrid clinical community is warmly acknowledged, especially that from Dr Ruth Warren, Iqbal Warsi and Jane Ding of Addenbrookes Hospital, Cambridge, UK, and Dr Chiara del Frate and Professor Massimo Bazzocchi of the Istituto di Radiologia at the Università degli Studi di Udine, Italy. Last, but by no means least, the authors are indebted to the former MammoGrid Project Coordinator, Roberto Amendolia, both for his pioneering contribution and for his robust support.